\documentstyle[aps,pra]{revtex}  
\begin{document}                
\draft
\title{\bf Entanglement between motional states of a single trapped ion and light}
\author{F.L. Semi\~ao, A. Vidiella-Barranco and J.A. Roversi} 
\address{Instituto de F\'\i sica ``Gleb Wataghin'',
Universidade Estadual de Campinas,
13083-970   Campinas  SP  Brazil}  
\date{\today}
\maketitle
\begin{abstract}    
We propose a generation method of Bell-type states involving  light and the 
vibrational motion of a single trapped ion. The trap itself is supposed to be 
placed inside a high-$Q$ cavity sustaining a single mode, quantized electromagnetic 
field. Entangled light-motional states may be readily generated if a conditional 
measurement of the ion's internal electronic state is made after an appropriate 
interaction time and a suitable preparation of the initial state. We show that all 
four Bell states may be generated using different motional sidebands (either 
blue or red), as well as adequate ionic relative phases.
\end{abstract}

The investigation of trapped ions manipulated by laser beams \cite{wine0} is of 
importance not only due to the fundamental physics involved, but also because of 
potential aplications, such as precision spectroscopy \cite{wine1} and 
quantum computation \cite{cira0}. The laser fields couple the (quantized) internal 
degrees of freedom in the ion to the (quantized) vibrational motion of the ion's 
center of mass, but the fields themselves are usually treated as classical. The 
quantization of the field of course brings new possibilities.
Within that realm, it has been already investigated the influence of the field 
statistics on the ion dynamics \cite{zeng,knight}, as well as the transfer of 
coherence between the motional states and light \cite{parkins}. There is much 
interest in the generation of non product, entangled states, and trapped ions seem 
to constitute a suitable system for doing that \cite{wine2}. Entangled states 
involving atoms rather than photons may also be used for testing Bell's inequality 
\cite{wine3,knight1}, as it has been already experimentally demonstrated \cite{wine3}. 
In general, what has been achieved so far is either the entanglement between the 
internal degrees of freedom of a single ion (electronic states) with the vibrational 
motion states of the ion itself, or the
entanglement between internal states of several ions \cite{wine2,solano}. 
Nevertheless, there are few discussions about possibilities of entanglement 
between the quantized field and the vibrational motion of the ion. This might be of 
special interest in quantum information; an entangled state of a subsystem that may
{\it store quantum information} (vibrational motion) with a subsystem that can be used 
for the {\it propagation of quantum information} (light). As another example of 
entanglement between matter and light, we may refer to a recently reported scheme for 
entangling light with atoms in a Bose-Einstein condensate \cite{meyst}.

In this contribution we present a simple scheme through which there 
could be produced entanglement between the (center of mass) vibrational motion of 
a single trapped ion and the electromagnetic field. We show that it is possible to 
generate the whole Bell state basis simply by choosing either the blue or the red
sideband (with different relative phases between ionic states).

We consider a single trapped ion, within a Paul trap, which is by its 
turn placed inside a high-$Q$ cavity \cite{blatt}, so that the cavity mode 
couples to the internal electronic states of the ion as well as to the vibrational 
degrees of freedom, as it has been already discussed in \cite{knight}. The 
hamiltonian corresponding to such a system may be written as 
\begin{equation}
\hat{H}=\hbar\nu \hat{a}^{\dagger}\hat{a} + \hbar\omega\hat{b}^{\dagger}\hat{b}
+\hbar\frac{\omega_0}{2}\sigma_z +\hbar g(\sigma_+ + \sigma_-)(\hat{b}^{\dagger}+
\hat{b})\sin\eta(\hat{a}^{\dagger}+\hat{a}).
\label{H}
\end{equation}
Here $\hat{a}^{\dagger}(\hat{a})$ denote the creation (annihilation) operators of
the center-of-mass vibrational motion of the ion (frequency $\nu$), 
$\hat{b}^{\dagger}(\hat{b})$ are the creation (annihilation) operators of photons 
in the field mode (frequency $\omega$), $\omega_0$ is the ionic vibration frequency, 
$g$ is the ion-field coupling constant, and $\eta=2\pi a_0/\lambda$ is the Lamb-Dicke
parameter, being $a_0$ the amplitude of the harmonic motion and $\lambda$ the 
wavelength of light. In the Lamb-Dicke regime, i.e., if the ion is confined in a 
region much smaller than light's wavelenght ($\eta\ll 1$), we may write
$\sin\eta(\hat{a}^{\dagger}+\hat{a})\approx\eta(\hat{a}^{\dagger}+\hat{a})$. This 
is of course a convenient way of linearizing the hamiltonian, although another 
approach, based on a unitary transformation of the hamiltonian and which avoids the 
application of Lamb-Dicke approximation from  the beginning, is also possible 
\cite{ours}. If we tune the light field to the first red sideband, 
i.e., $\delta=\omega_0-\omega=\nu$, we obtain, after discarding the rapidly 
oscillating terms, the following interaction picture hamiltonian:
\begin{equation}
\hat{H}^r_I= \eta\hbar g (\sigma_- \hat{a}^{\dagger} \hat{b}^{\dagger} +
\sigma_+ \hat{a} \hat{b}).
\label{HI}
\end{equation}
Such a hamiltonian describes the simultaneous process of creation (annihilation) of 
one quanta of vibrational motion, one quanta of the field, while the atom has its 
internal energy decreased (increased). The corresponding evolution operator 
$\hat{U}^r(t)=\exp(-i\hat{H}^r_I t/\hbar)$ will be, in the atomic basis
\begin{equation}
\hat{U}^r(t)=\hat{C}_{n+1} |e\rangle\langle e| + \hat{C}_{n} |g\rangle\langle g|
-i\hat{S}_{n+1} \hat{a}\hat{b} |e\rangle\langle g|
- i \hat{a}^\dagger\hat{b}^\dagger \hat{S}_{n+1} |g\rangle\langle e|,
\label{EO}
\end{equation}
where 
\begin{equation}
\hat{C}_{n+1}=\cos\left(\eta g\sqrt{(\hat{a}^\dagger\hat{a}+1)
(\hat{b}^\dagger\hat{b}+1)}\:t\right),
\end{equation}
\begin{equation}
\hat{C}_{n}=\cos\left(\eta g\sqrt{\hat{a}^\dagger\hat{a}\:
\hat{b}^\dagger\hat{b}}\:t\right),
\end{equation}
and
\begin{equation}
\hat{S}_{n+1}=\frac{\sin\left(\eta g\sqrt{(\hat{a}^\dagger\hat{a}+1)
(\hat{b}^\dagger\hat{b}+1)}\:t\right)}{\sqrt{(\hat{a}^\dagger\hat{a}+1)
(\hat{b}^\dagger\hat{b}+1)}}.
\end{equation}

We may now investigate the time evolution of the state vector having the following
initial condition for the ion-field state
\begin{equation}
|\Psi(0)\rangle= |n\rangle_f |m\rangle_v (\cos\theta |e\rangle + e^{i\phi}\sin\theta
|g\rangle),
\label{IC}
\end{equation}
or the field prepared in a number state $|n\rangle_f$ containing $n$ photons, the
ion's center of mass motion prepared in a number state $|m\rangle_v$ containing $m$ 
quanta, and the ion's internal levels prepared in a coherent superposition of two 
energy eigenstates  $|\varphi\rangle= \cos\theta |e\rangle + e^{i\phi}\sin\theta 
|g\rangle$. This particular initial condition is crucial for the generation of 
entanglement between the states of ionic vibration and the electromagnetic field. 
Such an initial superposition state in equation (\ref{IC}) may
be prepared through the convenient application of laser pulses. At a time 
$t$, the ion-field state vector will become 
\begin{eqnarray}
|\Psi(t)\rangle=\left[\cos\theta \cos\left(\eta g\sqrt{(n+1)(m+1)}\:t\right)|n\rangle_f
|m\rangle_v -ie^{i\phi}\sin\theta \sin\left(\eta g\sqrt{nm}\:t\right)|n-1\rangle_f
|m-1\rangle_v
\right]|e\rangle \\ \nonumber
+\left[e^{i\phi} \sin\theta \cos\left(\eta g\sqrt{nm}\:t\right)|n\rangle_f
|m\rangle_v -i\cos\theta\sin\left(\eta g\sqrt{(n+1)(m+1)}\:t\right)|n+1\rangle_f
|m+1\rangle_v
\right]|g\rangle.
\label{TE}
\end{eqnarray}
The resulting state above is an entangled state involving the ion's internal 
(electronic) degrees of freedom, the vibrational motion and the cavity field. If one 
measures the internal state of the ion (either in $|g\rangle$ or $|e\rangle$), that action
will collapse the state $|\Psi\rangle$ into entangled states of ionic vibrational 
motion and the cavity field.
For instance, we may consider that the ion's motion is initially cooled down to the 
vacuum state 
$|0\rangle_v$ and the cavity field is also in its vacuum state $|0\rangle_f$. In 
this case, for interaction times $t_k= \pi(4k+1)/2\eta g$ ($k=0,1,2,\ldots$) and for 
equally weigthed ionic states ($\theta=\pi/4$), if one measures (via fluorescence) 
the ion in its internal state $|g\rangle$, the resulting vibration-light field 
state will become
\begin{equation}
|\psi\rangle= \frac{1}{\sqrt{2}}\left(e^{i\phi}|0\rangle_f|0\rangle_v -i|1\rangle_f 
|1\rangle_v\right),
\label{BS}
\end{equation}
which is a Bell-type state, or an entangled state having as subsystems both the ion
and cavity field. Note that the ionic relative phase $\phi$ in the superposition 
of electronic states $|e\rangle$ and $|g\rangle$ is fully transferred to the 
resulting entangled state. Although there is no direct interaction between the 
motion of the trapped ion and the cavity field, the coupling of the ion's internal 
(electronic) levels to both vibration and light makes possible the generation of states 
of the type shown above in equation (\ref{BS}).

There is also the possibility of tuning the light field to the first blue sideband, 
or $\delta=-\nu$. The corresponding interaction hamiltonian will then read
 \begin{equation}
\hat{H}^b_I= \eta\hbar g (\sigma_- \hat{a} \hat{b}^{\dagger} +
\sigma_+ \hat{a}^{\dagger} \hat{b}).
\label{HIR}
\end{equation}
In this case, while the ion has its internal energy increased, a quanta of its 
vibrational motion is created and a photon is annihilated. We may follow a similar
procedure as we have done for the first red sideband, and find out what type of entangled 
states may be generated under such circumstances. The corresponding evolution operator
is 
\begin{equation}
\hat{U}^b(t)=\hat{C}'_{n+1} |e\rangle\langle e| + \hat{C}'_{n} |g\rangle\langle g|
-i\hat{S}'_{n} \hat{b}\hat{a}^\dagger |e\rangle\langle g|
- i\hat{b}^\dagger\hat{a} \hat{S}'_{n} |g\rangle\langle e|,
\label{EOB}
\end{equation}
where 
\begin{equation}
\hat{C}'_{n+1}=\cos\left(\eta g\sqrt{(\hat{b}^\dagger\hat{b}+1)
\hat{a}^\dagger\hat{a}}\:t\right),
\end{equation}
\begin{equation}
\hat{C}'_{n}=\cos\left(\eta g\sqrt{\hat{b}^\dagger\hat{b}
(\hat{a}^\dagger\hat{a}+1)}\:t\right),
\end{equation}
and
\begin{equation}
\hat{S}'_{n}=\frac{\sin\left(\eta g\sqrt{(\hat{b}^\dagger\hat{b}+1)
\hat{a}^\dagger\hat{a}}\:t\right)}{\sqrt{(\hat{b}^\dagger\hat{b}+1)
\hat{a}^\dagger\hat{a}}}.
\end{equation}
We may prepare the initial state as the one in equation (\ref{IC}), but with $n=0$ 
(field in the 
vacuum state) and $m=1$ (ion vibrational motion in the first excited state). For
interaction times $t_k$ (the same as to the red sideband case), after having 
detected the ion in the internal state $|g\rangle$, the resulting state will be 
\begin{equation}
|\psi'\rangle= \frac{1}{\sqrt{2}}\left(e^{i\phi}|0\rangle_f|1\rangle_v -i|1\rangle_f 
|0\rangle_v\right),
\label{GBSA}
\end{equation}
which is also a Bell-type state involving the quantized cavity field as well 
as the ion's vibrational motion. In fact, by taking $\phi\pm\pi/2$ 
in equations (\ref{GBSA}) and (\ref{BS}), we are able to obtain the four states 
cosntituting the {\it Bell state basis} 
\begin{equation}
|\psi\rangle_{\pm}= \frac{1}{\sqrt{2}}\left(|0\rangle_f|0\rangle_v \pm |1\rangle_f 
|1\rangle_v\right),
\label{BB1}
\end{equation}
and
\begin{equation}
|\psi'\rangle_{\pm}= \frac{1}{\sqrt{2}}\left(|0\rangle_f|1\rangle_v \pm |1\rangle_f 
|0\rangle_v\right).
\label{BB2}
\end{equation}

More general states could also be generated, depending on the initial conditions. For 
instance, in the red sideband case ($\delta=\omega_0-\omega=\nu$), if the ion is initially
prepared in a state having $m$ excitations, $|m\rangle$, and the field in the vacuum
state, the resulting entangled state will be (after measuring the ion in the internal
state $|g\rangle$)
\begin{equation}
|\psi\rangle= \frac{1}{\sqrt{2}}\left(e^{i\phi}|0\rangle_f|m\rangle_v -i|1\rangle_f 
|m+1\rangle_v\right).
\label{GBS}
\end{equation}

Note that entanglement here involves states belonging to different kinds of physical
systems (although both sub-spaces have infinite dimension), but because of their 
nature (light and massive trapped ions) new possibilities for quantum information 
processing might arise. Needless to say that the generation of such entangled states 
is of importance for addressing fundamental issues in quantum theory as well. 
We have so far considered an ideal situation in which losses are not taken into
account. Dissipation introduced by the finite-$Q$ cavity will of course become an
important origin for decoherence, together with losses related to the ion trap 
itself. Therefore the Bell state generation here proposed must be accomplished
within the cavity decay time, which is long enough in high-$Q$ cavities available
nowadays (around $0.2$ s) \cite{walther}. The loss of a photon will obviously  
destroy the Bell state generated in such a scheme. Nevertheless we shall remark that
losses, which are normally regarded as responsible for decoherence effects, might 
not only induce nonclassical behaviour \cite{mine} but also assist the generation of 
pure states \cite{knight2}. In the method presented in reference \cite{knight2}, 
an atom-atom entangled state is generated through a (properly monitored) photon 
decay. In our case, although both subsystems constituting the Bell states are 
vulnerable to losses, it would be possible to preserve their integrity for 
relatively long times \cite{walther}. 

In summary, we have proposed a generation scheme of entangled states of two 
coupled harmonic oscillators, the ionic vibrational motion and a cavity field. We have 
shown that it is possible to generate Bell-type states having rather simple initial state 
preparation, e.g., the vacuum state for both cavity field and the ion motion.

\acknowledgments

This work is partially supported by CNPq (Conselho Nacional para o 
Desenvolvimento Cient\'\i fico e Tecnol\'ogico), and FAPESP (Funda\c c\~ao 
de Amparo \`a Pesquisa do Estado de S\~ao Paulo), Brazil, and it is linked 
to the Optics and Photonics Research Center (FAPESP).

\end{document}